\documentstyle{l-aa}
\begin{document}

\def\etal{{\it et al.\ }}
\def\eg{{\it e.g.,}}
\def\ie{{\it i.e.,}}
\def\Halpha{H$\alpha$}
\def\Hbeta{H$\beta$}
\def\Hgamma{H$\gamma$}
\def\Hdelta{H$\delta$}
\def\Lya{Ly$\alpha$}
\def\Lyb{Ly$\beta$}



   \thesaurus{11         
              (11.19.3; 11.3.2; 11.9.5; 11.5.2;11.9.4)} 

\title{VCC 144: a star-bursting dwarf galaxy in the Virgo Cluster}


\author{N. Brosch \inst{1}, E. Almoznino \inst{1} \and G. Lyle Hoffman \inst{2}} 
\institute{The Wise Observatory and 
the School of Physics and Astronomy \\ Raymond and Beverly Sackler Faculty 
of Exact Sciences \\
Tel Aviv University, Tel Aviv 69978, Israel \and Department of Physics, Lafayette College 
\\Easton PA 18042, U.S.A.}

\thanks {Based on observations by the International
    Ultraviolet Explorer (IUE) collected at the Villafranca Satellite
    Tracking Station of the European Space Agency}

\offprints{N. Brosch}

\date {Received  ; accepted  }

\maketitle

\begin{abstract}
We describe results of a multi-spectral study of a blue compact dwarf galaxy in 
the Virgo Cluster. The object was observed with broad-band 
and H$\alpha$ imaging, ultraviolet observations, and radio synthesis. 
Our data were combined
with previously published optical observations, with HI single-beam observation 
and with far-infrared data, and were compared 
to results of evolutionary synthesis programs. The radio synthesis 
observations revealed a compact concentration of HI coincident with the optical galaxy, 
embedded in a diffuse, asymmetric HI cloud which has no trace of optical emission.
While the overall velocity dispersion and size of the HI structure 
suggests that the total mass in stars and gas is not sufficient to
gravitationally bind the system as a whole, the HI clump coincident 
with the optical galaxy requires little or no dark matter to be 
self-gravitating.  The diffuse cloud  has more 
complex velocity structure and is extended in a direction approximately
perpendicular to the optical major axis.
 
The optical-UV data  can be explained by a single population of stars formed in a 
recent burst, indicating that this is a genuine young galaxy.
The efficiency of star formation is similar to that in large disk galaxies.
The IR emission indicates the presence of dust; this must have
been formed very recently, or was already present within the original
HI cloud from which the galaxy was formed. The round and smooth isophotes, the
correspondence of the optical and HI redshifts, and
the lack of any suitable nearby galaxy, indicate that the starburst was probably not
triggered by an external interaction with a visible galaxy. The
distribution of HI fits better a blow-out scenario than an accretion or
collision with a companion.
We point out features in common with other actively star-forming dwarf galaxies   
and conclude that, at least in the southern outskirts
of the Virgo cluster, intensive star formation, perhaps for the
first time in some objects, takes place at present.


\end{abstract}


\section{Introduction}
Star formation (SF) is a fundamental process in the evolution of galaxies.
But this process is far from being
well understood. The SF is usually characterized by the
initial mass function (IMF) and the total SF rate (SFR),
which depends on many factors such as the density
of the interstellar gas, its morphology, its metallicity, {\it etc.}

According to Larson (1987), four major factors drive star 
formation in galaxies: large scale gravitational instabilities, cloud compression 
by density waves, compression in a rotating galactic disk due to shear
forces, and random cloud collisions. In galaxies with previous stellar
generations, additional SF triggers exist, such as shock waves from
stellar winds and supernova explosions.  
In dense environments, such as clusters of galaxies and
compact groups, tidal interactions and collisions
with other galaxies probably play some role in triggering star
formation ({\it e.g.,} the blue compact dwarf
and the Im galaxies apparently formed at the ends of the tidal tails
of Arp 105: Duc and Mirabel 1994).

While ``global''phenomena, such as the first two SF
triggers of Larson (1987), play a large part in grand design spirals,
random collisions of interstellar clouds
may provide the best explanation for dwarf galaxies with bursts of SF.
Therefore, understanding SF in dwarf galaxies should be
simpler than in other types of galaxies.
Understanding this type of objects has also
cosmological implications, as some star forming dwarf galaxies may
be genuinely young.

One way to derive the star formation history of 
galaxies is with a  combination of  credible spectral synthesis models
and observations in as many  spectral bands 
as possible, which should provide constraints on the IMF.
This approach is usually followed for a sample of 
galaxies of similar type, in which the star formation parameters would be
basically similar. Surface photometry in broad-bands, 
from which the spectral energy distribution (SED) is derived,
together with a tracer of massive stars, such as H$\alpha$ 
or UV measurements, 
makes it possible to obtain the total ongoing SFR.
The H$\alpha$ radiation is directly coupled
to the radiation at wavelengths $\lambda<$912\AA\, ({\it e.g.}, Kennicutt 
1983), indicating the ongoing star formation. UV radiation longward of the
Lyman break, mainly emission detected by space experiments such
as IUE and HST, which is longward of Lyman $\alpha$, is produced 
not only by the most massive stars, but also by cooler stars
that do not ionize hydrogen significantly.

 Almoznino (1995) studied a sample of BCDs in the Virgo Cluster selected from the 
Binggeli, Sandage, and Tamman (1985, hereafter BST) catalog, with 
HI measurements from Hoffman \etal 
(1987, 1989a). Among the objects in his sample he found one 
exceptional object, VCC 144, which could be 
a truly young galaxy. Here we describe our observations of this
object, combine them with existing information, compare them with
evolutionary population synthesis models, and present
our interpretation.

The Virgo cluster contains a very unusual object (HI 1225+01),
discovered serendipitously by Giovanelli \& Haynes (1989) as an extragalactic
HI cloud within the cluster. Subsequently, an optical counterpart
was identified ({\it e.g.,} Djorgovski 1990), was studied
 by Salzer \etal (1991), and was recently mapped in HI with the VLA
(Chengalur \etal 1995). Salzer \etal showed it to be
a system which witnessed a burst of star formation $\sim$1 Gyr ago, 
with signs of a recent ($\sim$10 Myr) second burst. Chengalur \etal (1995)
found rotation in the HI distribution and identified the star-forming clump
as the only region where the HI surface density exceeds 4.6 10$^{20}$ atoms
cm$^{-2}$. The system is probably the result of a tidal interaction between two
HI components. 

A general review of properties of dwarf galaxies was given by Thuan (1992) and a
review of spectroscopic properties of star-forming dwarfs was presented by
V\'{i}lchez (1995).
Lately, a number of papers reported extended HI envelopes, or HI companions, of 
dwarf galaxies ({\it e.g.,} Taylor \etal 1995; Van Zee \etal 1995;
 Hoffman \etal 1996; Szomoru \etal 1996a, 1996b). As will be seen below, we detected an 
extended HI envelope connected with VCC 144.
It is  possible that this object represents a more intensely star-forming case 
than HI 1225+01 and that dwarf galaxies with extended HI distributions have
other common characteristics. Therefore, we will compare our measurements and derived
properties for this Virgo BCD with those of the optical counterparts of the
dwarf galaxies with extended HI and with HI 1225+01.

\section{Observations}

Below we first review published observations of VCC 144 and describe our
own observations together with additional data collected from the literature or data banks.
We adopt here a uniform distance of 18 Mpc to the Virgo cluster, and by inference
to all its members, including VCC 144. A smaller cluster distance, 
$\sim14$ Mpc, is suggested by the [OIII] distances to planetary 
nebulae (Jacoby \etal 1990), but Visvanathan \& Griersmith (1979) 
give 17.9 Mpc, in agreement with Sandage \& Tammann (1974). We note other 
values for the distance to the Virgo cluster, in particular 16.8 Mpc (Tully 1988)
and the estimate of the distance to M100 from observations of Cepheids
(17.1 Mpc: Freedman \etal 1995), while considering the depth of the cluster to be
some 3-4 Mpc.
Note that VCC 144 could be located in the W cloud (Hoffman \etal 1989b). If
this is the case, its distance could be 1.7 to 2 times greater than adopted
here.

\subsection{Optical and UV observations}
 
BST identified VCC 144 [optical peak at $\alpha$(J2000)=12$^h$ 15$^m$ 18$^s$.35,
 $\delta$(J2000)=+05$^{\circ} $45' 39".2], a BCD galaxy, as a 
member of the Virgo cluster, despite it being just outside the borders
of the W cloud. Note that the coordinates given in BST are off the peak of
the optical image by $\Delta \alpha$=--0$^s$.65 and $\Delta \delta$=4".7.
de Vaucouleurs \etal (1991) quote for this galaxy
an optical heliocentric velocity v$_{\odot}$=1960$\pm$52 km s$^{-1}$.
  Gallagher \& Hunter (1986) reported on
single-aperture photoelectric photometry of this object, along with 64 other
similar objects. They measured within a 19" aperture 
 V=14.87$\pm$0.03, (B--V)=0.38$\pm$0.03, and (U--B)=--0.51$\pm$0.03.

Gallagher \& Hunter (1989, GH89) surveyed  spectroscopically
BCDs in the Virgo Cluster. For VCC 144 they measured an H$\beta$ equivalent
width EW[H$\beta$]=35\AA.
The low resolution ($\sim$15\AA\,) spectrum with the IRS on the
KPNO 0.9 m telescope, shown as their Fig. 2, exhibits a blue-sloping continuum 
with strong and narrow emission
lines. These are (in order of strength) [OIII] $\lambda$5007, [OII] $\lambda$3727,
[OIII] $\lambda$4959, H$\beta$, H$\gamma$, H$\delta$ (very weak),
and a blend longward of [OII], that can be mostly HeI $\lambda$3888\AA\,
or an incomplete sky subtraction.
The region between H$\beta$ and H$\gamma$ is free
of the HeII $\lambda$4686\AA\, and NIII $\lambda$4641\AA\, high
excitation lines.
GH89 correct upward the H$\beta$ line by 10\% to account for underlying absorption
(which is not evident in their spectrum), and obtain
a line flux of 1.14 10$^{-13}$ erg cm$^{-2}$ s$^{-1}$, with ratios of other
lines to H$\beta$: [OII] $\lambda$3727=3.16, [OIII] $\lambda$4959=1.03,
and [OIII] $\lambda$5007=3.68. To these we add, by measuring from their Fig. 2:
[OIII] $\lambda$4363\AA\,$\leq$0.07, H$\gamma\simeq$0.4, H$\delta\simeq$0.1,
 He I $\lambda$3888\AA\,$\simeq$0.3 (if real). 

VCC 144 was observed with the CCD camera of the Wise Observatory (WiseObs)
on April 11, 1991.
The galaxy was imaged through standard B, V, R and I filters. In 
total, we collected three B images each of 10 minute exposure, three 200 sec
V-band images, and four 120 sec exposures for R and for I with the RCA-CCD. 
This is a thinned, back-illuminated chip, with 512$\times$320 pixels. Each
pixel subtends 0".87 at the f/7 focus of the 1m WiseObs reflector.
A reference star close enough to appear on the galaxy's CCD
frame was imaged together with the galaxy. 
This star was calibrated against Landolt standards (Landolt 1973, 1992)
with photoelectric photometry at the WiseObs using a 30" round aperture.
The galaxy is shown as the combined B-band frame, underlying the contours
in Fig.~\ref{Fig4}.

The galaxy was also imaged through narrow-band filters, to
derive the H$\alpha$ emission. Two filters
were used: one containing the H$\alpha$ line at the redshift
of the object and the other sampling the continuum at $\lambda>\lambda(H\alpha$).
 Three 20 min exposures were obtained through each of the two filters.
In addition, the HZ44 spectrophotometric standard star
was measured several times each night through different air
masses with the same filters, in order to derive the
atmospheric extinction and the absolute photometric calibration for the 
H$\alpha$ images.

%
EA observed VCC 144 with IUE in 19 Jan 93  for a full VILSPA observing shift
with $\sim$6 hours of integration. The spectrum was
 obtained through the IUE large aperture with the short wavelength 
primary camera in low dispersion setup. VCC 144 is
a very compact object, and the radiation comes almost entirely from
its central 10", thus IUE collected essentially all of its UV flux.
The IUE spectrum (SWP 46761) was very weak after its extraction.
We measured two regions 100\AA\, wide and averaged the flux densities
within each to obtain two UV photometric indices at 1350\AA\, and at
1850\AA\,. The monochromatic magnitudes in these two bands are
[1350]=13.50$\pm$0.16 and [1850]=14.10$\pm$0.22 and the respective
flux densities are 1.35$\pm$0.20 10$^{-14}$ and 0.78$\pm$0.16 10$^{-14}$ 
erg s$^{-1}$ cm$^{-2}$ \AA\,$^{-1}$.

\subsection{IRAS and radio data}

VCC 144 is not listed in the
IRAS point source catalog. The IRAS
data consist of flux densities in four infra-red (IR) bands centered at
12$\;\mu m$, 25$\;\mu m$, 60$\;\mu m$ and 100$\;\mu m$. IRAS scanned 
most of the celestial sphere several times, and it is possible to co-add
the various IRAS scans for a given location to obtain a deeper detection or upper
limit at this location in the sky. The IRAS Faint Source Catalog lists a 
source very close to the position of VCC 144 with the following flux
densities: [12]=0.150$\pm$0.023 Jy, [25]=0.312$\pm$0.047 Jy, [60]=0.622$\pm$0.093
Jy, and [100]=0.657$\pm$0.099 Jy. We adopt this as the FIR emission from the
galaxy.

As mentioned above, our sample is based on the HI measurements of Hoffman {\it et al.}
(1987) and Hoffman {\it et al.} (1989) with the Arecibo radio telescope.
They list for VCC 144 a flux integral of 2305$\pm$19 mJy km s$^{-1}$ and a 135 
km s$^{-1}$ HI line width at 20\% of the peak. The
profile is triangular-shaped, asymmetric, and shows no sign of a 
two-horn profile expected for a flat rotation curve.  While a solid-body
rotation curve can give rise to a triangular profile (Skillman 1996), it is unusual to find
evidence of significant rotation in dwarf galaxies which exhibit triangular
central beam profiles as narrow as this one (Hoffman \etal 1996).
The widths at 50\% and 80\% of the peak
intensity are 82 and 44 km s$^{-1}$. The heliocentric velocity from the HI measurement,
adopted by Hoffman \etal (1989) as the mid-point of the HI profile edges
at 50\% peak intensity, is 2014 km s$^{-1}$. Within the measuring accuracy,
the optical and HI redshifts coincide.
 
VCC 144 was observed in HI and radio continuum at the Very Large Array (VLA)
\footnote{The Very Large Array is part of the National Radio Astronomy Observatory 
which is operated by Associated Universities, Inc., under cooperative agreement 
with the U.S. National Science Foundation.}. Observations in the
L band continuum and HI line   were conducted for 15 and 82 minutes on-source, 
respectively, on 12 and 16 April 1988 in the C array; C band continuum observations 
were made for 52 minutes on-source on 10 September 1988 in the D array.
The same pointing center (1950), $12^h12^m45.0^s + 06^{\circ} 02' 
20"$, was used throughout.
The spectral line observations used 32 channels with channel 
spacing of 10.45 km ${\rm s}^{-1}$ centered on heliocentric velocity 
2014 km ${\rm s}^{-1}$.
Continuum observations were conducted with 50 MHz bandpasses at 
6 and 20 cm. Calibration was performed using sources from the VLA calibrator 
list in the usual way, with map-making and deconvolution performed using the standard 
routine IMAGR within the NRAO AIPS software package.
For the HI data, the continuum was subtracted in the $uv$ domain, and zero-spacing 
fluxes were interpolated for each channel from the Arecibo spectrum.
	The rms noise achieved after CLEANing was 0.21 mJy/beam for
	the L band continuum data, 0.053 mJy/beam for the C band
	data, and 2.0 mJy/beam in each channel for the HI line data.


\section{Results}

We measured for VCC 144 the following integrated photometric
properties: V=14.83$\pm$0.03, (B--V)=0.46$\pm$0.03, (V--R)=0.22$\pm$0.02,
and (R--I)=0.23$\pm$0.02. The photometry was performed through a polygonal
aperture, which traced the ``limits'' of the galaxy as estimated
visually on the deepest (V-band) image. This corresponds to a
surface brightness of $\sim$25.5 mag/square arcsec. This rather
shallow limit is the result of the low exposure collected for VCC 144,
which is one of the brightest galaxies in our sample.

\begin{figure}
     \picplace{7cm}
\includegraphics{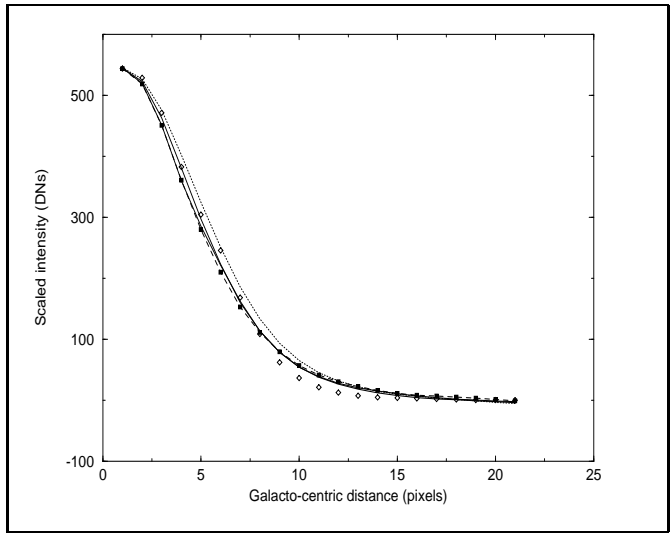}
    \caption{Intensity profiles of the net, sky subtracted, images of VCC 144.
The profiles have been normalized to the peak of the B-band profile, to ease
the shape comparison. The different profiles are as follows: B-heavy solid line,
V-light solid line, R-dashed line with filled squares, I-dotted line, 
H$\alpha$-open diamonds.}%
         \label{Fig1}%
    \end{figure}

Although the V magnitude we measure reproduces
that by GH86, there is a possible discrepancy in the B band, as the
color (B--V) is marginally redder (2$\sigma$) than what they measured. The 
difference could be related to a radial color gradient in the galaxy,
as GH86 measured only the innermost 19" while we include an additional
$\sim$10" external region. 
However, a test of the image profiles in the B, V, and R
bands (Fig.~\ref{Fig1}, below) shows that they are $\sim$identical, thus 
VCC 144 is {\it not}
redder in its outer regions. With the measured photometry, and
an adopted distance of 18 Mpc to the Virgo cluster, VCC 144 appears to be
a low luminosity dwarf galaxy (M$_V\simeq$--16.4). If in Cloud W, as mentioned
above and $\sim1.7\times$ further away, M$_V\simeq$--17.6 mag, it would still
be a dwarf galaxy, although on the brighter side.


The H$\alpha$ frames were combined and the final image was shifted 
to match the broad-band images, and the galaxy was measured inside the same 
polygonal area used for its broad-band image. The method of measurement was 
basically the same as for the broad-band frames,
with the exception  that the sky background level was computed several 
times using different 'boxes' around the objects (see Almoznino \etal 1993). 
 The mean of these results
was adopted as the object counts, while the error was taken as
the largest of the errors of each measurement. The standard deviation of
the results was smaller than this. 
The integrated H$\alpha$ flux was measured to be 7.05$\pm$0.76 10$^{-13}$
erg s$^{-1}$ cm$^{-2}$. The average line equivalent width,
obtained by dividing the net line image by the off-line continuum
image, is 159$\pm$22\AA\,, with a peak value of 390\AA\,.
The intense H$\alpha$ emission implies that $\sim$13\%  of the R band
photons originate from this line and from [NII] emission.
We have not corrected the magnitudes and colors for line emission, either
for H$\alpha$+[NII] in R, for [OIII] in V, or for [OII] in U. The contribution
of line emission is strongest in the R band; the other lines have smaller
equivalent width and are located near the edges of bands.

We smoothed all final images, combined from different images through 
the same filter, with a gaussian of full width at half-maximum of 4 pixels=3".5,
which corresponds to the typical seeing in our images. We then obtained an 
intensity profile for each image, by fitting ellipses to the isophotes. The 
intensity profiles for the B, V, R, and I bands, as well as that for the net-H$\alpha$
image, were scaled to the same peak intensity (of the B image), and the
widths of the profiles were compared. We found that all five images had 
virtually identical profiles, to within 5\% of the peak intensity at 
$\sim$12 pixels$\simeq$11" galactocentric distance. The intensity profiles 
are plotted in Fig.~\ref{Fig1}. The images are all resolved, with very similar 
eccentricities of $\sim$0.6 in all colors. The surface brightness is high with
an average value of $\sigma_B$=20.5 mag/$\Box$". Note that because of the smoothing,
there are only very few ``independent'' points, thus fitting the light distribution
to any of the usual profiles seems pointless.

The high surface brightness, the strong H$\alpha$ emission, and the 
compact and blue appearance of the galaxy 
 point to an interpretation favoring a strong burst of star formation
taking place at present in VCC 144. We did not find any faint extensions 
or disturbances in our images; thus the triggering of star formation 
in VCC 144 cannot be attributed to a recent interaction with another 
visible galaxy.

\begin{figure}
\picplace{8 cm}
\includegraphics{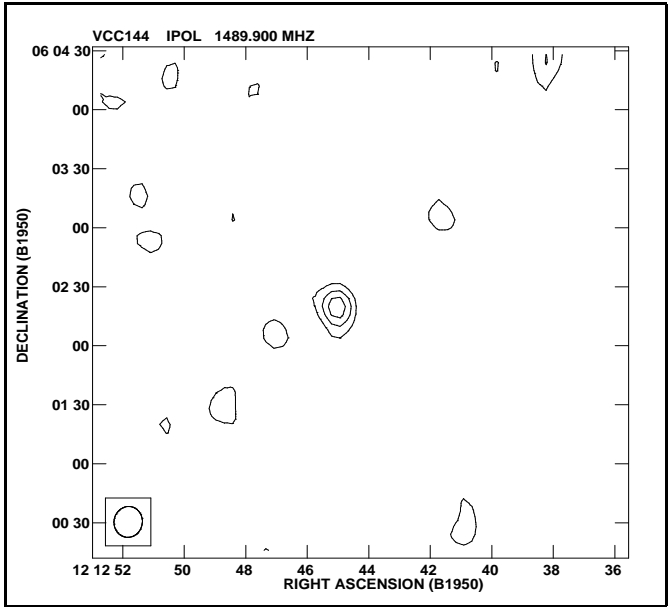}
\caption{20 cm VLA continuum map of VCC 144. Contours are 0.5, 1.0, 
1.5, 2.0 and 2.5 mJy/beam.  The beam is shown at lower left.
Only the single Gaussian feature at the center of the map appears to be 
related to VCC 144.}
\label{Fig2}
\end{figure}

\begin{figure}
\picplace{8 cm}
\includegraphics{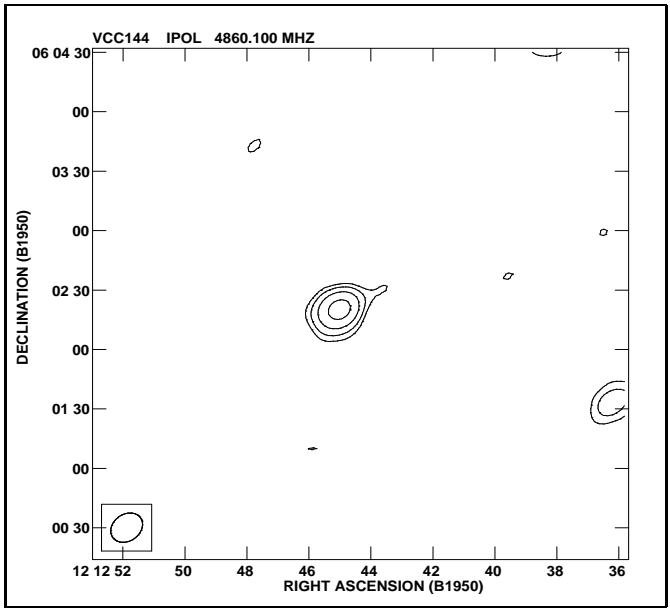}
\caption{ 6 cm VLA continuum map. 
Contours are 0.15, 0.30, 0.60, and 1.2 mJy/beam.
The beam is shown at lower left.
Only the single Gaussian feature at the
center of the map appears to be related to VCC 144.}
\label{Fig3}
\end{figure}

Both VLA continuum images (Figs.~\ref{Fig2} and~\ref{Fig3}) show a single Gaussian 
feature centered on the optical image of the galaxy.
At 20 cm, the feature has a size $9.6" \times 3.0"$ 
after deconvolution with the $15.7" \times 13.7"$ beam.
The peak intensity of a fitted Gaussian is $1.93 \pm 0.31$ mJy/beam and 
the integrated intensity over the feature is $2.35 \pm 0.65$ mJy.
At 6 cm the galaxy is still only marginally resolved
(the other features on the map are all well outside the optical
and HI boundaries, and are almost certainly unrelated to VCC 144),
with a size about $5.0" \times 
2.9"$ after deconvolution with the $16.7" \times 13.2"$ beam, a Gaussian peak at 
$1.66 \pm 0.10$ mJy/beam and an integrated intensity of 1.78$\pm$0.19 mJy.
This gives a global spectral index of 0.23$\pm$0.25 if the integrated intensities 
are used, or 0.13$\pm$0.15 if the peak intensities are used instead.

The 21 cm line data were processed at full resolution (16") and at 45",
to bring out extended features at lower column densities. The maps at full 
and at reduced resolution are presented below and are discussed in detail.

\begin{figure}
\picplace{8 cm}
\includegraphics{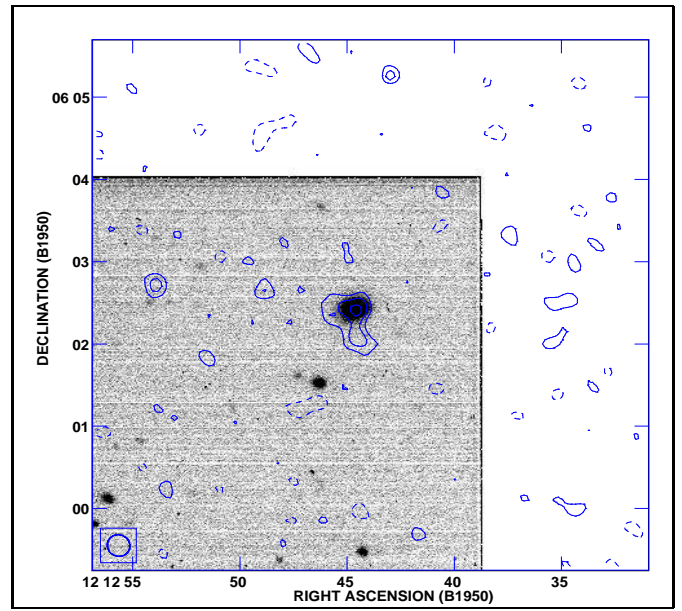}
\caption{ Total HI map of VCC 144 at 16" resolution. 
The HI contours are overlaid on a grey scale plot of the B image. 
Contours are drawn at $-8.0$, 8.0, 12, and 16 $10^{20}$ atoms cm$^{-2}$.}
\label{Fig4}
\end{figure}

The total HI map at 16" resolution was computed by integrating over the 
emission-bearing channels with no blanking applied.
The resulting contour map, overlaid on a grey scale version of
the B-band image, is shown in Fig.~\ref{Fig4} and reveals one concentration of about 
$1.2 \times 10^8\;M_{\odot}$ coincident in position with the optical 
image, with hints of more diffuse emission extending perhaps one
beam width to the southwest, and about 1.5 arcmin to the northeast.

\begin{figure}
\picplace{8 cm}
\includegraphics{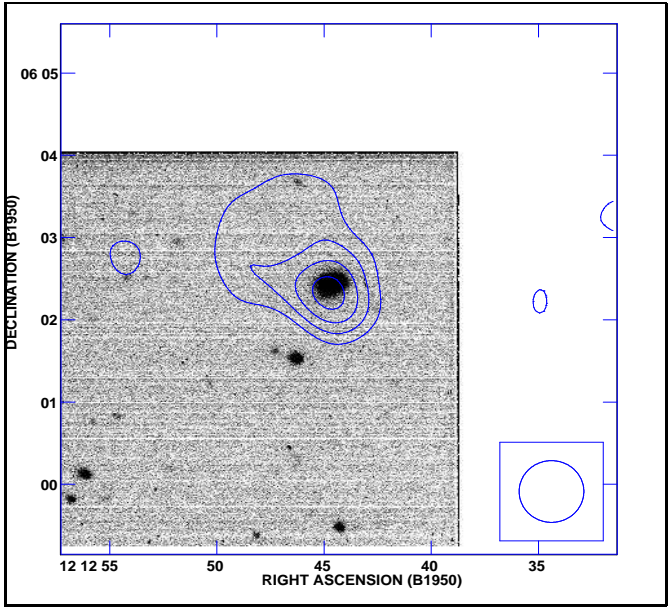}
\caption{ Total HI map of VCC 144 at 45" resolution obtained by convolving 
the 16" resolution data cube with a Gaussian overlaid on the blue image. 
Contours are drawn at 2.5, 3.8, 5.1, and 6.3 $10^{20}$ atoms cm$^{-2}$.}
\label{Fig5}
\end{figure}

To confirm that the extension is real, we convolved the map with a Gaussian 
to smooth it to a resolution of 45 arcsec.
The result is shown in Fig.~\ref{Fig5}, and the NE extension stands out clearly.
In fact, the two figures show that the HI distribution extends on both sides
of the minor axis of the galaxy.

Integrating over the entire HI distribution gives a total mass 
of $2.7 \times 10^8\;M_{\odot}$, 
about 50\% larger than the flux estimated from the single-beam spectrum 
obtained with the Arecibo telescope, which partially resolves the HI
cloud.
Our images and inspection of the Palomar Sky Survey copies show no 
optical counterpart within the NE extension to a surface brightness limit 
of $\sim$24 mag/square arcsec. The HI linear extent is about 11.5 kpc.
All masses are appropriate to a distance of 18 Mpc as assumed
throughout, and would be larger by a factor of $(D / 18 {\rm Mpc} )^2$
if VCC 144 were at the W cloud distance.

\begin{figure}
\picplace{8 cm}
\includegraphics{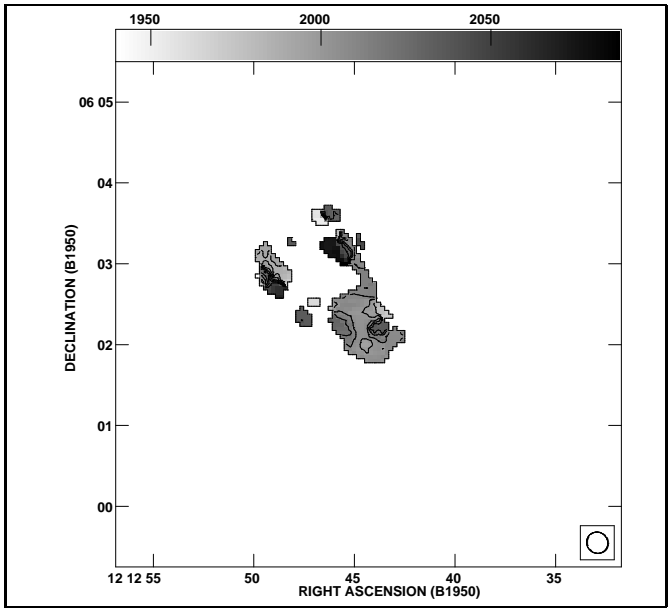}
\caption{ Isovelocity contour map of VCC 144 at 16" resolution 
superimposed on a grey scale image of the same.
The grey scale is indicated at the top of the figure; contours are shown 
at 1990, 2000, 2010, 2020, 2030, and 2040 km s$^{-1}$.
The beam is shown at lower right corner.}
\label{Fig6}
\end{figure}

To obtain isovelocity contours at 16" resolution it was necessary to blank the 
cube in two stages:  first everything outside the lowest contour on the smoothed 
total HI map (Fig.~\ref{Fig5}), presumably unrelated to VCC 144, was blanked.
Then only features in each channel that exceeded 2 mJy/beam were retained for 
computation of the first moment map, shown in Fig.~\ref{Fig6}.

\begin{figure}
\picplace{8 cm}
\includegraphics{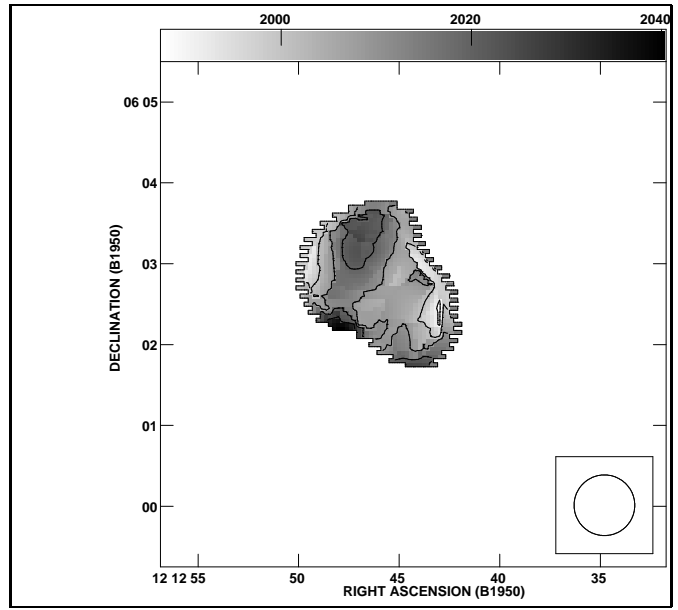}
\caption{The same as Fig.~\ref{Fig6}, but for 45" resolution.
The same grey scale and contours were used. The beam is shown in the lower right corner.}
\label{Fig7}
\end{figure}

\begin{figure}
\picplace{9 cm}
\includegraphics{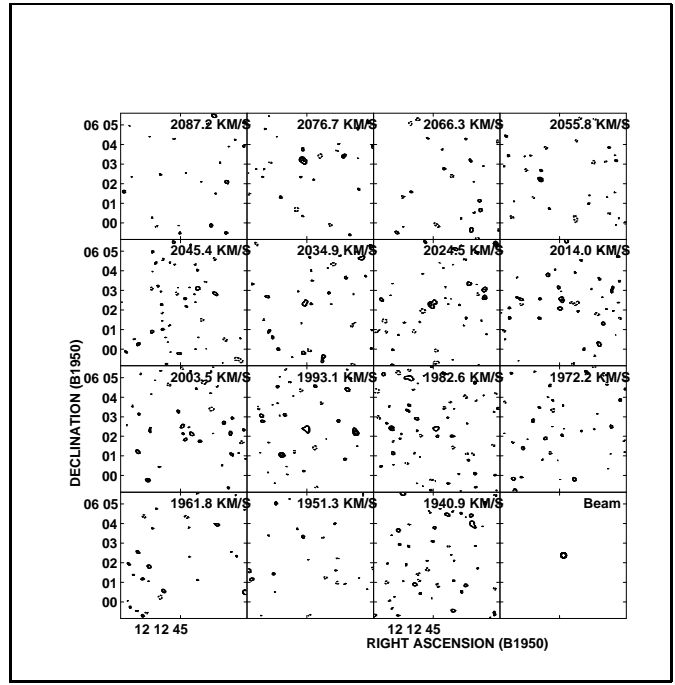}
\caption{ Mosaic of channel maps of VCC 144 at 16" resolution. 
The central (heliocentric) velocity of each channel labels it on the upper right corner.
Contours are drawn at $-5$, 5, 7.5, 10.0 and 12.5 mJy/beam.
The beam is indicated in the lower right panel.}
\label{Fig8}
\end{figure}

A 45" resolution first moment map was also produced, retaining only features 
in each channel of the spatially smoothed cube that exceeded 5 mJy/beam.
Everything outside the lowest contour on the smoothed total HI map was then 
blanked from the smoothed first moment map, which is shown in Fig.~\ref{Fig7}. 

\begin{figure}
\picplace{8 cm}
\includegraphics{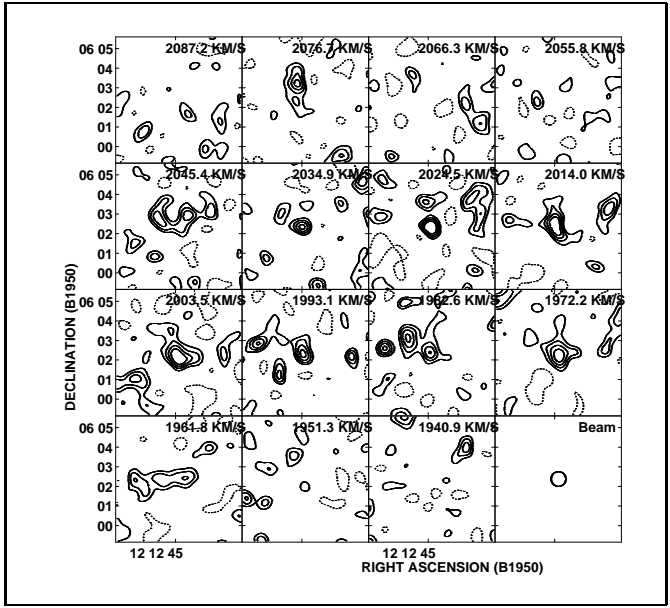}
\caption{The same as Fig.~\ref{Fig8}, but for 45" resolution. 
The central (heliocentric) velocity labels each channel in the upper right corner.
Contours are drawn at $-5$, 5, 7.5, 10.0 and 12.5 mJy/beam.
The beam is indicated in the lower right panel.}
\label{Fig9}
\end{figure}

Inspection of the isovelocity contours in Figs.~\ref{Fig6} and~\ref{Fig7}, 
and the individual channel maps shown in Figs.~\ref{Fig8} and~\ref{Fig9} as
mosaics, leads 
us to the following interpretation:  the
dominant clump, centered on the optical image, has a mean velocity around 2010 
km ${\rm s}^{-1}$ with a roughly Gaussian velocity profile of width about 
80 km ${\rm s}^{-1}$. The NE extension lies generally at higher velocity, but has a
more complex velocity profile with prominent narrow features at about
2070, 2040 and 1980 km ${\rm s}^{-1}$.
This suggests some bulk motion of the diffuse gas with respect 
to the main concentration, but we cannot say 
 conclusively from these observations
	whether that motion is more properly described as infall,
	ejection, or rotational motion.  Other considerations, to
	be discussed below, lead us to favor a gas blow-out.
We do not have adequate signal nor spatial resolution to determine whether 
the main concentration rotates about its own axis, or has purely turbulent 
internal motions. In any case, turbulent motions appear to be at least as significant as 
ordered motions in the dynamics of the clump and of the system as a whole.

\section{Discussion}

\subsection{Total mass and masses of non-stellar components}

We estimate here a representative total dynamical mass for the system, and 
masses of its different components (stars, gas, and dust). Dwarf irregular galaxies
often show little rotation and their dynamics are dominated by turbulent
motions ({\it e.g.,} Sargent \etal 1983).
It is possible to estimate an indicative gravitational mass M$_G$ for
the galaxy using the half-width at half-intensity of the single beam 21 cm profile 
$\Delta v$:
\begin{equation}
M_G=\frac{(\Delta v)^2 R}{G}
\end{equation}
For a typical HI maximal extent from the center of the HI distribution of R=4 kpc
the gravitational mass is M$_G$=1.56 10$^{9}$ M$_{\odot}$. 

The total B magnitude measured by us, with the adopted distance to the galaxy,
yields a total blue luminosity of $\sim$3.6 10$^8$ L$_{\odot}$, making the
indicative mass-to-blue light ratio  M$_G$/L$_B\approx$4.2. This is 
somewhat higher than M/L values of blue star-forming galaxies. Note though 
that if we are witnessing a collision between two clouds, and if the relative
velocities of the clouds are approximately perpendicular to our line of
sight, there is no requirement for the gas to be bound to the system. In
this case we might be seeing a tidal disruption taking place, or the
blow-out of gas from a system experiencing strong star formation.

The indicative dynamical mass of the dominant clump, assuming a profile
half-width of 40 km s$^{-1}$ and a radius of 1.6 kpc, is $\sim$5.7 10$^8$
M$_{\odot}$. For the densest part of the NE extension (visible as a distinct 
clump in the 16" resolution total HI map), taking ``profile half-width'' to be
half the difference between the velocities of the two narrow peaks and a
similar radius, we get 3.6 10$^8$ M$_{\odot}$.

We calculated the HI content of VCC 144 from the HI flux density integrated over 
the entire map, following the 
prescription in de Vaucouleurs {\it et al.} (1991).  The flux integral
value from Hoffman \etal (1989a) yields a total
HI mass of 1.8$\pm$0.4 10$^8$ $(\frac{D}{18Mpc})^2$ M$_{\odot}$, and the total 
integrated flux from the VLA map yields 2.7 10$^8$ $(\frac{D}{18Mpc})^2$ M$_{\odot}$ 
with similar uncertainty.
The VLA map of the dominant clump, which coincides with the optical image, 
contains 1.2 10$^8$ 
M$_{\odot}$ of HI. The HI mass is thus comparable with the stellar mass, as
derived from the blue luminosity for a M/L$\approx$1 assumption, but is
only $\sim$20\% of the indicative dynamical mass.

The location of VCC 144 in a FIR color-color diagram, such as that by Helou (1986),
coincides with that of starburst, BCD, and extragalactic HII region galaxies,
and the exact location of the point indicates that most FIR emission is produced 
by the current star formation,
{\it i.e.,} there is no significant ``cirrus'' cloud contribution.
The total dust content can be estimated from the relations of  Thronson \etal (1988),
Bothun \etal (1989),  or Thuan \& Sauvage (1992).
The  harmonic mean of the three values yields M$_d$=9.3 10$^3$ M$_{\odot}$ 
as representative total dust mass and
 the dust-to-gas mass ratio is $\sim$3.4 10$^{-5}$ if the dust is distributed within
the entire HI entity or $\sim$7.8 10$^{-5}$ if it is relegated to the
optical galaxy and its HI cloud. Some 10$^4$ massive stars
must have existed prior to the present star formation burst to have produced
this amount of dust, or the dust could be pre-stellar in this galaxy.

\begin{figure}
\picplace{7 cm}
\includegraphics{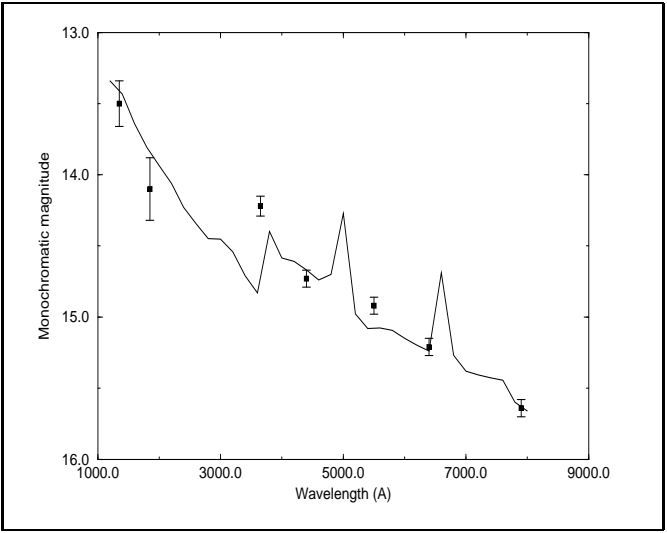}
\caption{Spectral energy distribution of VCC 144. The values derived from our
IUE and WiseObs observations are shown as monochromatic magnitudes with
error bars. The continuous line is the spectral energy distribution, in monochromatic
magnitudes, for a young starburst with essentially no extinction (Kinney \etal 1996).}
\label{Fig10}
\end{figure}
 
It is useful to consider the entire SED
of the galaxy, displayed in Fig.~\ref{Fig10}. The points originate mainly from our
WiseObs broad-band photometry and from measurements with the IUE. In order 
to derive the total U magnitude we assumed
that the (U--B) color measured by GH86 for the inner 19" is representative for the
entire galaxy. The SED shows a steep decline
with increasing wavelength, both in the UV and in the optical. Since our derived
SED is very sparse because of having only broad-band colors,
it cannot be tested against detailed spectra, but can be
compared, in general, to various typical SEDs of different stellar populations.
A comparison with the experimentally-derived spectral templates of Kinney
\etal (1996) in Fig.~\ref{Fig10} shows a good fit to the ``negligible extinction'' 
[E(B--V)$\leq$0.10] template of a starburst (their Fig. 2g). Thus, from 
the SED alone, we could 
conclude that VCC 144 {\it is} an unextinguished, star-bursting galaxy.
  
The ``recent'' star formation rate may be estimated from the blue
luminosity (Gallagher \etal 1984, GHT) as:
\begin{equation}
SFR_B=6.5 \, 10^{-9} L_B=2.34 M_{\odot}/yr
\end{equation}
  averaged over the past 0.4--6 Gyrs.
The assumption in GHT is that the SFR is constant over
the lifetime of the galaxy and the IMF is a Salpeter law with an
upper mass cutoff of 100 M$_{\odot}$. A recent starburst would have correspondingly
more blue and luminous stars, lowering the SFR resulting from the same L$_B$.

The galaxy shows the most intense H$\alpha$ emission among objects in our sample. 
The line has very high equivalent width, peaking at 390\AA\,. The entire $\sim$1 
kpc optical galaxy is a huge HII region. The line radiation comes from the entire galaxy 
and, as mentioned above, no optical extension to the galaxy is visible
beyond the line emitting region even in the broad-band images.
This intense H$\alpha$ emission indicates a high SFR at present.
The lower limit on the [OIII] line ratio from the Gallagher \& Hunter (1989)
spectrum indicates an electron temperature $\leq13,000$K.

The present star formation rate is estimated from the H$\alpha$, again using a
relation from GHT:
\begin{equation}
SFR_{H\alpha}=1.27 \, 10^9 \, F(H\alpha) \times D^2
\end{equation}
This requires correcting the H$\alpha$ emission for extinction.
Adopting our H$\alpha$ flux and comparing it with the H$\beta$ from GH89,
we find H$\alpha$/H$\beta$=6.18, apparently indicating a significant amount of extinction
for Case B recombination theory. The comparison is valid, although our H$\alpha$ 
measurement is {\it global} while that of GH89 was taken through a 22" aperture,
because the H$\alpha$ image of VCC 144's is $\sim 25$ pixels$\simeq$22" wide,
but is elongated. GH89
estimate E(B--V)$\approx$0.0 from H$\beta$/H$\gamma$. The H$\alpha$/H$\beta$ 
inconsistency,
produced by two different observations, may result if GH89 missed part of the 
total H$\beta$ flux in their round aperture. We conclude that the 
extinction for VCC 144 is probably ``negligible'', as indicated
by the reasonable fit to the Kinney \etal (1996) SED.
This yields SFR$_{H\alpha}$=0.29 M$_{\odot}$/yr, assuming no extinction 
of the H$\alpha$ line.

The discrepancy between SFR$_B$ and SFR$_{H\alpha}$ found above may result from
an incompatibility of the model. A similar result was found for an irregular galaxy
by Thronson \etal (1988). As will be shown below, our preferred interpretation
of VCC 144 is of a recent and first star formation burst. Therefore, the GHT
method of deriving SFR$_B$ will not be correct. A comparison with the spectral
energy distribution of different young starbursts  may be more 
adequate.  The lack of HeII and typical WR emission signatures indicates that high 
mass stars (M$_*\geq$40M$_{\odot}$)
are missing from the present population mix, while GHT assumed an upper mass
cutoff of 100 M$_{\odot}$; only if the SF burst happened instantaneously a few 10$^6$
yrs ago, the most massive stars and their WR remnants could be absent now, 
which would explain the discrepancy between SFR$_B$ and SFR$_{H\alpha}$.
Another hint about the present population of massive stars, derived from the
radio continuum mapping, will be discussed below.

The far-IR flux, from \eg \, Lonsdale \etal (1985), is:
\begin{equation}
F_{FIR}=1.26 \, [2.58 \, S_{60} + S_{100}] \, 10^{-14}
\end{equation}
For VCC 144, at the nominal distance of 18 Mpc, this implies a FIR luminosity
of 7.6 10$^7$ L$_{\odot}$, thus $\frac{L_B}{L_{FIR}}$=4.7 and the galaxy is
not a strong FIR source.
The SFR in VCC 144 can also be estimated from the IRAS data, using the
method of Thronson and Telesco (1986). This assumes 
that essentially the entire luminosity of the massive OB stars is absorbed by 
the dust and is reemitted in the infrared. This yields the following relation:
\begin{equation} 
SFR_{FIR} = 6.5\times 10^{-10} L_{IR} 
\end{equation}
with the FIR luminosity $L_{IR}$ given in solar units and approximated by:
\begin{equation} 
L_{IR} \approx 6\times 10^5 D^2 (2.58 f_{60} + f_{100}) 
\end{equation}
where $D$ is the distance to the object in Mpc and $f_{60}$ and $f_{100}$
are the two IRAS flux densities in Jy. The SFR resulting from this method is
similar to that derived from the H$\alpha$ data ($\sim$0.3 M$_{\odot}$ yr$^{-1}$), 
increasing our confidence in the ``negligible extinction'' case for this galaxy
and, coincidentally, in the assumption that most photons from OB stars
heat the dust.

It is worthwhile to consider a distance-free parameter, such as the star formation
per unit galactic area. This can be derived using the size of the galaxy from
its CCD image, or profile, as exhibited in Fig. 1.
In this case, the ongoing SFR per unit area is given by:
\begin{equation} 
 SFR/area = 1.078\times10^8 F(H\alpha)/\Box"
\end{equation}
where the SFR/area is in M$_{\odot}$/yr/$pc^2$ and 
$F(H\alpha)/\Box"$
is the observed H$\alpha$ flux per square arcsecond. The high SFR/area value,
2.28$\times10^{-7}M_{\odot}/yr/pc^2$, originates from the strong luminosity and 
compactness of the object. Not surprisingly, with a high specific SF but small size
VCC 144 undergoes strong star formation per unit
galactic area, but shows only modest total SFR. For comparison, the Milky
Way has a total SFR of $\sim 5 M_{\odot}/yr$ and SFR/area of $\sim 7\times 10^{-9}
M_{\odot}/yr/pc^2$. Large, active, star forming galaxies have SFR/area
similar to that of VCC 144 ({\it e.g.,} Pogge and Eskridge 1987).  
Thus, once the star formation activity is triggered, its
intensity, as manifested by the SFR/area, is similar in small and large galaxies.

This interpretation implies that mechanisms believed to be responsible for star
formation in large galaxies but not in dwarf galaxies, such as compression due to 
density waves or from rotational shear forces, are as efficient
as other mechanisms, such as random collision between clouds or gravitational
instabilities caused by other effects. Otherwise, there would be significantly
more star formation per area in large star-forming galaxies than in late-type dwarfs.
It is worth mentioning that here the {\it efficiency} of the star formation
process, caused by the various mechanisms, is tested, rather than how wide-spread 
these mechanisms are. Therefore, the conclusion at this stage is
that once star formation is induced, its efficiency is approximately 
the same, regardless of the mechanism which induced it.

The radio synthesis observations indicate that the continuum source is  
connected with the presence of stars, as the emission is localized in the cloud
which coincides with
the optical counterpart. The spectral index is consistent with thermal emission
processes. The thermal radio continuum can be calculated from the physical 
conditions in the HII region:
\begin{equation}
\frac{S_\nu}{1 Jy}=1.33 \, 10^{-47} N_C (\frac{\nu}{1 GHz})^{-0.1} (\frac{T_e}{10^4 K})^{0.45} (\frac{D}{1 kpc})^{-2}
\end{equation}
For the observations at 20 cm, with T$_e$=1.3 10$^4$K (upper limit, from the [OIII] lines),
with the number of ionizing photons calculated from H$\alpha$ (see below) and at 18 Mpc, 
S(1420 MHz)=0.9 mJy. At 5 GHz the thermal
radio continuum is 0.8 mJy and makes up $\sim$30\% of the detected flux.
The fraction of non-thermal 20 cm contribution is $\sim$1.5 mJy and at 6 
cm it is $\sim$1.0 mJy. The resultant non-thermal spectral index is $\sim$0.3.
With an average flux from a SNR at 18 Mpc of $\sim$1 mJy there should be at present at most
a couple of radio SNRs in VCC 144. This is consistent with the present number of early-type
stars; with $\sim$2000 O7 stars present (see below), with a lifetime to supernova explosion
of $\sim$10$^7$ years and a radio SNR lifetime of 10$^4$ yrs, a
couple of SNRs should be detectable at present in radio continuum.

\subsection{Time scales and comparison with stellar evolution models }

Above we derived a number of SFRs, which can be combined with the observationally-derived
masses to yield two indicative time scales. In particular, it is
possible to use the ``star formation timescale'',  defined by Hodge (1993) as:
\begin{equation}
\tau_{SF}=\frac{M_{stars}}{SFR_0}
\end{equation}
where $M_{stars}$ is the total mass in stars, estimated with the
assumption of $\frac{M_{stars}}{L_B}=1$,
 and $SFR_0$ is the present SFR. For VCC 144 we obtain
$\tau_{SF}$=1.2 Gyrs, similar to Hodge's value for the most extreme starburst
galaxy in his list. This is an indication on the production of the present-day luminosity
with solar-type stars; for an IMF rich in young stars  M/L is lower. Mateo (1992) found
 one Magellanic cluster with M/L$\approx$0.1 for an age of 8 10$^7$ yrs.
 For three Magellanic clusters, Elson and Freeman (1985) found M/L between 0.11 and 0.56;
if this is the case for VCC 144, the mass in stars would be lower and
$\tau_{SF}$ could be  $\sim10^8$ yrs or even less.

The Roberts time scale measures the time to exhaust the galactic
HI reservoir given the present SFR:
\begin{equation}
\tau_R=\frac{M(HI)}{SFR_0}
\end{equation}
We obtain for VCC 144 $\tau_R$=0.9 Gyrs if the entire HI reservoir is
considered, or 0.4 Gyrs for the HI coincident with the optical image. 
The estimate of the Roberts time scale
assumes no recycling of material from the stars; otherwise, the time to
exhaust the HI will be longer for a constant SFR. For a star formation proceeding
in bursts followed by long periods of quiescence, the galaxy could survive 
for a Hubble time, provided the star formation duty cycle is $\approx$2\%.
This constraint can be relaxed if the total HI would include the diffuse extended cloud of hydrogen.


We compared our results with population synthesis models of Leitherer \& Heckman 
(1995) and found reasonable matches for single star burst models 5-7 Myrs
old. We emphasize that the matches are not perfect, but indicative.
We tested also the H$\alpha$ data through the number of ionizing photons, 
following Osterbrock (1989) for a simple Case B recombination theory 
and with the H$\alpha$ luminosity in erg/s:
\begin{equation}
N_c=7.43 \; 10^{11} \times L(H\alpha)
\end{equation}
The color index (H$\alpha$--V) for the models, for this purpose, is obtained as
\begin{equation}
[H\alpha-V]=129.8-2.5 \, log(N_c)-M_V
\end{equation}
We compared the observed color index [H$\alpha$--V]=--2.5 log [F(H$\alpha$)]--V=15.56, 
(B--V), and M$_V$ with the model predictions. The indices fit a $\sim$6 Myr old
single star burst model from Leitherer \& Heckman (1995) but are not compatible
with continuous star formation.
These are much too blue to fit the observations.
From the value of N$_c$ found for VCC 144, using 
L(H$\alpha$)=2.7 10$^{40}$ erg s$^{-1}$ and adopting a Lyman continuum photon flux
of 10$^{49}$ s$^{-1}$ from a typical O star, it appears that  
some 2000 late-O or B stars are needed in VCC 144 at present to ionize the hydrogen.
For a Salpeter IMF, this implies that the present star burst produced
some 10$^6$ M$_{\odot}$ in stars of all masses. 

The observational and derived parameters related to VCC 144 are summarized in the 
table below. For comparison, we included a few ``typical'' values from the
compilation of V\'{i}lchez (1995, V) for star-forming dwarfs, from Huchra (1977, H) for
Markarian galaxies, and from Thuan (1992, T) for late-type dwarfs.
The metallicity was derived from the [OII]+[OIII] lines as measured from GH89
data and it compares well with the typical metallicity found by V\'{i}lchez (1995)
for his sample of dwarf galaxies. We conclude that our observations, combined with
data from the literature, indicate that a single, short and recent burst of star
formation can account for the observed properties of this object.

\begin{table}
\begin{tabular}{ccc}
Observed properties  & VCC 144 & SF DG \\ \hline
V & 14.83 \\
B--V & 0.46 & 0. 4 (V) \\
U--B  & --0.51 & --0.36 (H) \\
V--R  & 0.22 & 0.08--1.17 (H) \\
R--I  & 0.23 \\
UV--V  & --1.33\\
IRAS ? &  Weak \\
d$_{maj}\times$d$_{min}$ &   $\sim$24"$\times$12" \\
Velocity [km s$^{-1}$]:  Optical   & 1960$\pm$52 \\
Velocity [km s$^{-1}$]: Radio &   2014 \\
Total H$\alpha$ flux [erg s$^{-1}$ cm$^{-2}$] &   7.05$\pm$0.76  10$^{-13}$ \\
([OII]+[OIII])/H$\beta$ &  7.78 \\
HI flux integral [Jy km s$^{-1}$] &   2.31 \\ \\
Derived properties & VCC 144 & SF DG  \\ \hline
d [kpc] &   $\sim$1 & $<$0.5--20.6 (V) \\
L$_B$ [L$_{\odot}$]  & 3.6  10$^8$ \\
L(H$_{\alpha}$) [ erg s$^{-1}$] &   2.7  10$^{40}$ & 10$^{38}$--10$^{41}$ (V)\\
N(O7) &   2000 \\
M(HI) [M$_{\odot}$] &   1.8  10$^8$ & 10$^8$ (T) \\
M(HI)/L$_B$ [M$_{\odot}$/L$_{\odot}$] &   2.0 & 2, 4 (T) \\
M(dust) [M$_{\odot}$] &  9.3 10$^3$ \\
12+log(O/H) &   $<$8.2 or 7.7 & 7.7--8.9 (V)\\
M$_G$ [M$_{\odot}$] &  1.6  10$^{11}$ \\
SFR [M$_{\odot}$ yr$^{-1}$]  & 0.3 \\
log(SFR/area) [M$_{\odot}$ yr$^{-1}$ pc$^{-2}$] &   --6.64 \\
log $\tau_R$ [yrs] &  8.78 \\
\end{tabular}
\end{table}

\subsection{Comparison with similar objects}

We mentioned in the introduction the unique object HI 1225+01, a faint blue 
dwarf galaxy with large amounts of hydrogen and a companion HI cloud, which are
probably interacting (Chengalur \etal 1995). HI 1225+01 is only $\sim$1.6 Mpc
away from VCC 144 (in projected distance) and both are located in the same 
general region of the Virgo cluster. Other similar objects are some of the
HII galaxies with HI companions mapped by Taylor \etal (1995, 1996), and some
of the BCDs found by van Zee \etal (1995) to have extended HI
envelopes. Even the arch-type young galaxy I Zw 18 shows an extended HI
envelope (Skillman \etal 1996). However, none are identical
with VCC 144.

In many aspects, the optical counterparts of HI 1225+01 and I Zw 18 are 
similar to VCC 144. The objects are blue and can be understood as  recent starbursts. 
Salzer \etal (1991) identified signs of an older starburst in HI 1225+01, which
enriched it with nitrogen and now produces at most $\sim$30\% of the B light.
Chengalur \etal (1995) concluded that the present burst of activity in 
HI 1225+01 was produced by a 
recent tidal interaction with the second HI lobe. Thus, its star forming activity 
could be explained, despite the object being relatively isolated. VCC 144 appears 
to be similarly isolated from other optical galaxies
at the southern outskirts of the Virgo cluster, with NGC 4197 at $\sim$54 kpc
projected distance as nearest neighbor, but shows a much more intense star 
forming activity.  

The BCDs studied by van Zee \etal (1995) are sometimes extended in HI
(16/41$\approx$40\% of the cases). An extended ($\sim3\times$ 
size of the optical galaxy), but symmetrical, envelope was found
for the nearby irregular galaxy Leo I (Young \& Lo 1996). In most cases
of diffuse HI near BCDs the HI extension is $\sim$symmetrical
around the optical object. However, in VCC 144 the diffuse HI appears in two 
extensions off the optical axis. As explained above, the iso-density contours of HI
shown in Figs.~\ref{Fig4} and ~\ref{Fig5} appear as two lobes on the 
extension of the galaxy
minor axis. The extensions, mainly the large one to the NE, represent large
amounts of kinetic energy, just from considering the mass and velocity
difference with respect to the HI on the optical counterpart ($\sim$5
10$^{52}$ erg). At least an equal amount of mechanical energy was probably
injected as turbulent motion in the blown-out gas. Leitherer \& Heckman (1995)
calculated a total mechanical energy injected by supernovae and stellar
winds of $\sim$1.6 10$^{55}$ erg (scaled to the present M$_V$ and for an
instantaneous starburst 6 10$^6$ yrs old). This indicates that a scenario of 
ISM blow-out from VCC 144, as a result of a recent starburst event, is a 
viable proposition.

\section{Conclusions}

\begin{enumerate}
 \item Optical observations of VCC144 show a blue compact, elliptical, bright dwarf
galaxy. The strong H$\alpha$ emission coincides and overlaps the broad band optical image of
the galaxy. 

 \item HI mapping with the VLA shows a main concentration coincident with the
optical galaxy and an asymmetric, diffuse extension which has no optical counterpart,
and which is extended along the minor axis of the galaxy. 
Rather large mass is required for the structure to be gravitationally confined.

 \item Comparisons with evolutionary synthesis models indicate that a most
probable explanation is of a first burst of star formation in the
last 10$^7$ years. The object does, however, contain dust.

 \item Similarities with other dwarf galaxies with extended HI envelopes are noted. 
We argue that a blow-out scenario is more appropriate for VCC 144 to explain
the shape of the HI envelope, than either
accretion, or a collision with an HI companion. The conclusion
is that galaxy formation, in the form of dwarf galaxies, takes place at present
in the Southern outskirts of the Virgo cluster.

\end{enumerate}

 \begin{acknowledgements}

Observations at the Wise Observatory are partly supported by a grant from the 
Israel  Science Foundation.
UV studies at the Wise Observatory are supported by special grants from
the Ministry of Science and Arts, through the Israel Space Agency, to
develop TAUVEX, a UV space imaging experiment, and by the Austrian Friends
of Tel Aviv University. GLH was supported in part by US National 
Science Foundation grant AST-9316213 to Lafayette College.
NB acknowledges the hospitality of Prab Gondhalekar
and of the IRAS Postmission Analysis Group at RAL, as well as IRAS Faint
Source catalog searches by Rob Assendorp. We are grateful to the referee, Daniel 
Kunth, for constructive remarks which improved the quality of this paper.

\end{acknowledgements}





\begin{thebibliography}{}

 \bibitem{}  Almoznino, E. 1995 PhD thesis, Tel Aviv University.

 \bibitem{}  Almoznino, E., Loinger, F. \& Brosch, N. 1993 MNRAS { 265}, 641.

 \bibitem{}  Bingelli, B., Sandage, A., \& Tammann, G.A. 1985 A.J. {  90}, 1681 (BST).


 \bibitem{}  Bothun, G.D., Lonsdale, C.J. \& Rice, W. 1989 ApJ {  341}, 129.


 \bibitem{}  Chengalur, J.N., Giovanelli, R. \& Haynes, M.P. 1995 AJ {  109}, 2415.


 \bibitem{}  Davidson, K. \& Kinman, T.D. 1985 ApJS {  58}, 321.

 \bibitem{}  de Vaucouleurs, G., de Vaucouleurs, A., Corwin, H.G., Buta, R.J., Paturel, G.,
\& Fouqu\'{e}, P. 1991 {\it The Third Reference Catalog of Bright Galaxies},
New York: Springer.

 \bibitem{}  Djorgovski, S. 1990 AJ {  99}, 31.


 \bibitem{}  Duc, P.-A. \& Mirabel, I.F. 1994 A\&A {  289}, 83.

 \bibitem{}  Elson, R.A.W. \& Freeman, K.C. 1985 ApJ {  288}, 521.


 \bibitem{}  Freedman, W.L., Madore, B.F., Mould, J.R., Hill, R., Ferrarese, L.,
Kennicutt, R.C., Saha, A., Stetson, P.B., Graham, J.A., Ford, H.,
Hoessel, J.G., Huchra, J., Hughes, S.M. \& Illingworth, G.D.
 1994 Nature {  371}, 757.
 
 \bibitem{}  Gallagher, J.S. \& Hunter, D.A. 1986 AJ {  92}, 557.

 \bibitem{}  Gallagher, J.S. \& Hunter, D.A. 1989 AJ {  98}, 806.

 \bibitem{}  Gallagher, J.S. Hunter, D.A. \& Tutukov, A.V. 1984 ApJ {  284}, 544.

 \bibitem{}  Giovanelli, R. \& Haynes, M.P. 1989 ApJ {  346}, L5.


 \bibitem{}  Helou, G. 1986 ApJ {  311}, L33.

 \bibitem{}  Hodge, P. 1993 in {\it Star Formation, Galaxies and the Interstellar
Medium} (J. Franco, F. Ferrini, and G. Tenorio-Tagle, eds.), Cambridge: Cambridge 
University Press, p. 294.

 \bibitem{}  Hoffman, G.L., Helou, G., Salpeter, E.E., Glosson, J. \& Sandage, A. 1987
ApJS {  63}, 247.

 \bibitem{}  Hoffman, G.L., Helou, G. Salpeter, E.E. and Lewis, B.M.
1989b ApJ {  339}, 812.

 \bibitem{}  Hoffman, G.L., Salpeter, E.E., Farhat, B., Roos, T., Williams, H. \&
Helou, G. 1996 ApJS {  105}, 269.

 \bibitem{}  Hoffman, G.L., Williams, H.L., Salpeter, E.E., Sandage, A. \& Binggeli, B. 
1989a ApJS {  71}, 701.

\bibitem{} Huchra, J.P. 1977 ApJS 35, 171.


 \bibitem{}  Jablonka, P. \& Arimoto, N. 1992 in {\it The Stellar Populations of Galaxies},
(B. Barbuy and A. Renzini, eds.), D. Reidel Publishing Company, p. 435.

 \bibitem{}  Jacoby, G.H., Ciardullo, R and Ford, H. 1990 ApJ {  356}, 332.

 \bibitem{}  Kennicutt, R.C. 1983 ApJ {  272}, 54.

 \bibitem{}  Kennicutt, R.C. 1989 in {\it The Interstellar Medium in Galaxies}
(H.A. Thronson \& J.M. Shull, eds.) Dordrect: Kluwer, p. 405.


 \bibitem{}  Landolt, A. 1973 AJ {  78}, 959.

 \bibitem{}  Landolt, A. 1992 AJ {  104}, 340.

 \bibitem{}  Larson, R.B. 1987 in {\it Starbursts and galaxy evolution} (T.X. Thuan,
T. Montmerle and J.T.T. Van, eds.),  Gif sur Yvette: Editions Frontieres.

 \bibitem{}  Leitherer, C. \& Heckman, T.M. 1995 ApJS {  96}, 9.

 \bibitem{}  Lonsdale, C.J., Helou, G., Good, J.C. \& Rice, W. 1985 {\it Cataloged
Galaxies and Quasars Observed in the IRAS Survey}, Pasadena: JPL.


 \bibitem{}  Mateo, M. 1992  in {\it The Stellar Populations of Galaxies},
(B. Barbuy and A. Renzini, eds.), D. Reidel Publishing Company, p. 147.

 \bibitem{}  Salzer, J.J., di Serego Alighieri, S., Matteucci, F., Giovanelli, R. 
\& Haynes, M.P. 1991 AJ {  101}, 1258.

 \bibitem{}  Sandage, A. and Tammann, G.A. 1974 ApJ {  194}, 559.

 \bibitem{}  Sargent, W.L.W., Sancisi, R. \& Lo, K.Y. 1983 ApJ {  265}, 711.

 \bibitem{}  Skillman, E.D. 1996, in {\it The Minnesota lectures on extragalactic
 neutral hydrogen} ASP Conf. Ser. 106 (E.D. Skillman, ed.), p.208


 \bibitem{}  Skillman, E.D., Palmer, R.C., Garnett, D.R. \& Dufour, R.J. 1996
in {\it From Stars to Galaxies}, ASP Conf. Ser. (C. Leitherer \etal, eds.), p. 366.

 \bibitem{}  Szomoru, A., van Gorkom, J.H. \& Gregg, M.D.
 1996a AJ {  111}, 2141.

 \bibitem{}  Szomoru, A., van Gorkom, J.H., Gregg, M.D. \& Strauss, M.A.
 1996b AJ {  111},  2150.

 \bibitem{}  Taylor, C.L., Brinks, E. \& Skillman, E.D. 1993 AJ {  105}, 128.


 \bibitem{}  Taylor, C.L., Brinks, E., Grashuis, R.M. \& Skillman, E.D. 1995 ApJS {  99}, 427
(erratum 1996 ApJS {  102}, 189).

 \bibitem{}  Taylor, C.L., Thomas, D.L., Brinks, E. \& Skillman, E.D. 1996 ApJS {  107}, 143.


 \bibitem{}  Thronson, H., Hunter, D.A., Telesco, C., Greenhouse, M. \& Harper,
D. 1988 ApJ {  334}, 605.

 \bibitem{}  Thronson, H.A., Jr. \& Telesco, C.M. 1986 ApJ {  311}, 98.
 
 \bibitem{} Thuan, T.X. 1992 in {\it Physics of nearby galaxies}
 (T.X. Thuan, Ch. Balkowsky, and J.T.T. Van, eds.) Gif-sur-Yvette: Editions
 Frontieres, p. 225.

 \bibitem{}  Thuan, T.X. \& Sauvage, M. 1992 A\&AS {  92}, 749.

 \bibitem{}  Tully, R.B. 1988 {\it Nearby Galaxies Catalog}, Cambridge: Cambridge
University Press.

 \bibitem{}  Pogge, R.W. \& Eskridge, P.B. 1987 AJ {  93}, 291. 

 \bibitem{}  Van Zee, L., Haynes, M.P. \& Giovanelli, R. 1996 AJ {  109}, 990.
 
 \bibitem{} V\'{i}lchez, J.M. 1995 AJ 110, 1090.

 \bibitem{}  Visvanathan, N. and Griersmith, D. 1979 ApJ {  230}, 1.

 \bibitem{}  Young, L.M. \& Lo, K.Y. 1996 ApJ {  462}, 203.

\end{thebibliography}
\end{document}